\documentclass[letterpaper,10pt]{proc}

\usepackage{eulervm}
\usepackage[margin=.5in]{geometry}

\makeatletter
\ifx\beamer@currentmode\undefined
  \usepackage[table]{xcolor}
\fi
\makeatother

\usepackage[utf8]{inputenc}
\usepackage{comment}
\usepackage{amsmath,amssymb,stmaryrd,textcomp}
\allowdisplaybreaks 
\usepackage[normalem]{ulem}
\usepackage{pgfpages}
\usepackage{utf8math}
\usepackage{url}
\usepackage{xspace}
\usepackage{graphicx}
\usepackage[hidelinks]{hyperref}
\usepackage{rcs}
\usepackage{enumitem}
\usepackage{makecell}

\newcommand{\HAX}{\text{H\kern-.1em A\kern-.1em C\kern-.05em S}\xspace}

\newcommand{\plank}{\ensuremath{\hslash}\xspace}

\newcommand{\CRSX}{\text{CRSX}\xspace}
\newcommand{\ie}{\textit{i.e.}\xspace}

\newcommand{\ov}[1]{\ensuremath{\mkern2mu\overline{\mkern-1.8mu{#1}\mkern-.8mu}\mkern1mu}}

\ifx\theorem\undefined
  \usepackage{amsthm}
  \swapnumbers 
  \newtheorem{theorem}{Theorem}[section]
  
  \theoremstyle{definition}
  \newtheorem{definition}[theorem]{Definition}
  \newtheorem{example}[theorem]{Example}
  
\fi
\ifx\remark\undefined
  \theoremstyle{definition}

  \newtheorem{remark}[theorem]{Remark}
\fi

\usepackage{alltt}
\usepackage{fancyvrb}
\CustomVerbatimEnvironment{code}{Verbatim}{tabsize=4,fontsize=\small,numberblanklines=false,xleftmargin=\parindent}
\CustomVerbatimCommand{\inputcode}{VerbatimInput}{tabsize=4,fontsize=\footnotesize,numbers=left,numberblanklines=false,xleftmargin=\parindent}

\usepackage{listings}
\lstnewenvironment{Java}[1][]{\lstset{language=Java,%
  basicstyle=\normalfont,%
  extendedchars=true,inputencoding=utf8,sensitive,%
  commentstyle=\color{blue}\sl,%
  identifierstyle=\textit,%
  keywordstyle=\kw,%
  basewidth={.5em,.4em},columns=flexible,numberstyle=\scriptsize,numberblanklines=false,xrightmargin=1pc,%
  #1}\upshape}{}

{\makeatletter \catcode`!=6 \catcode`\#=11 \catcode`\$=11 \catcode`\_=11 %
\gdef\justh@sh{\lst@um# }
\gdef\just@t{\lst@um @ }
\gdef\justd@llar{\lst@um$ }
\gdef\just@nder{\lst@um_ }
\gdef\Identifier!1{\expandafter\Identifier@two\the\lst@token \lst@um_ @}
\gdef\Identifier@two!1!2\lst@um_!3@{{%
  \def\1{!1}\def\2{!2}\def\3{!3}%
  \ifx\3\empty\else
    \expandafter\Identifier@subscript\3@%
  \fi
  \ifx\1\justh@sh              \mv{\expandafter\def\csname lst@um-\endcsname{\text{-}}\2\3}
  \else\ifx\1\justd@llar
    \ifx\2\empty               \kw{\$\3}%
    \else                      \kw{\expandafter\def\csname lst@um-\endcsname{\text{-}}\2\3}%
    \fi
  \else\ifx\1\just@at
    \kw{\expandafter\def\csname lst@um-\endcsname{\text{-}}\1\2\3}%
  \else\ifx\1\just@nder
    \con{\expandafter\def\csname lst@um-\endcsname{\text{-}}\1\2\3}%
  \else
    \edef\4{`\1=\noexpand\the\noexpand\uccode`\1}%
    \expandafter\ifnum\4\relax \con{\expandafter\def\csname lst@um-\endcsname{\text{-}}\1\2\3}%
    \else                      \var{\expandafter\def\csname lst@um-\endcsname{\text{-}}\1\2\3}%
  \fi\fi\fi\fi\fi
}}
\gdef\Identifier@subscript!1\lst@um_ @{\def\3{\ensuremath{\sb{\textup{!1}}}}}
}
\newcommand{\mv}[1]{\textsc{#1}}
\newcommand{\kw}[1]{\text{\bfseries\sffamily\color{black!60}#1}}
\newcommand{\kwm}[1]{\textcolor{black!60}{\ensuremath{\pmb{#1}}}}
\newcommand{\cw}[1]{\textsf{\color{gray}#1}}
\newcommand{\con}[1]{\textsf{#1}}
\newcommand{\var}[1]{\text{\sffamily\itshape#1}}

\lstdefinelanguage{HACS}{basicstyle=\sf,%
  extendedchars=true,inputencoding=utf8,sensitive,%
  identifierstyle=\Identifier,%
  string=[d]{"},morestring=[d]{'},upquote=true,showstringspaces=false,%
  comment=[l]{//},morecomment=[s]{/*}{*/},commentstyle=\color{blue}\rm,%
  keywordstyle=\kw,%
  keywords={as,attribute,data,fragment,global,import,main,module,nested,property,rule,scheme,simplify,sort,%
    space,static,sugar,symbol,tag,token,default,error,binds,template,variable},%
  literate=%
    {⟦}{{\ensuremath{\llbracket}}}1 {⟧}{{\ensuremath{\rrbracket}}}1 {⟨}{{\ensuremath{\langle}}}1 {⟩}{{\ensuremath{\rangle}}}1 %
    {[}{{\ensuremath{[\mkern.5mu}}}1 {]}{{\ensuremath{\mkern.5mu]}}}1 {⟨}{{\ensuremath{\langle}}}1 {⟩}{{\ensuremath{\rangle}}}1 %
    {→}{{\ensuremath{\rightarrow}}}2 {↑}{{\ensuremath{\uparrow}}}1 {↓}{{\ensuremath{\downarrow}}}1 {¬}{{\ensuremath{\lnot}}}2 {¶}{{\P}}1 %
    {ε}{{\ensuremath{\varepsilon}}}1 {…}{{\ensuremath{\dots}}}2 {∧}{{\ensuremath{\!\wedge}}}2 {∨}{{\ensuremath{\!\vee}}}2 {↦}{{\ensuremath{\mapsto}}}2 %
    {α}{{\ensuremath{\alpha}}}1 {β}{{\ensuremath{\beta}}}1 {γ}{{\ensuremath{\gamma}}}1 %
    {λ}{{\ensuremath{\lambda}}}1 {Λ}{{\ensuremath{\Lambda}}}1 %
    {γ}{{\ensuremath{\gamma}}}1 {Γ}{{\ensuremath{\Gamma}}}1 {ρ}{{\ensuremath{\rho}}}1 %
    {σ}{{\ensuremath{\sigma}}}1 {Σ}{{\ensuremath{\Sigma}}}1 {π}{{\ensuremath{\pi}}}1 {Π}{{\ensuremath{\Pi}}}1 %
    {λ̅}{{\ensuremath{\overline{\lambda}}}}1 {‾}{{\ensuremath{\overline{\phantom{x}}}}}1 %
    {¹}{{\ensuremath{^1}}}1 {²}{{\ensuremath{^2}}}2 {⌷}{{\ensuremath{\Box}}}1 %
    {@}{{\cw{@}}}1 %
    {@0}{{\cw{@0}}}2 {@1}{{\cw{@1}}}2 {@2}{{\cw{@2}}}2 {@3}{{\cw{@3}}}2 {@4}{{\cw{@4}}}2 %
    {@5}{{\cw{@5}}}2 {@6}{{\cw{@6}}}2 {@7}{{\cw{@7}}}2 {@8}{{\cw{@8}}}2 {@9}{{\cw{@9}}}2 %
}[comments,strings,keywords]
\lstnewenvironment{hacs}[1][]{\lstset{language=HACS,basewidth={.53em,.43em},columns=flexible,numberstyle=\scriptsize,numberblanklines=false,firstnumber=auto,xrightmargin=1pc,#1}\upshape}{}

\lstMakeShortInline[language=HACS,columns=fullflexible]"

\newcommand{\op}[1]{\ensuremath{\operatorname{#1}}}

\usepackage[all]{xy}

\makeatletter
\def\compacttableofcontents{{\bigskip
    \noindent\textbf{Contents:}~%
    \def\protect{\relax}%
    \def\contentsline##1{\csname contentslineA##1\endcsname}%
    \def\numberline##1{\DN@{##1}\ifx\next@\empty\else##1.~\fi}%
    \def\fresh##1{\relax \let\fresh=\nextfresh}%
    \def\nextfresh##1{\relax##1 }%
    \def\contentslineAsection##1##2##3{\fresh,##1~(\hyper@linkstart{link}{##3}{##2}\hyper@linkend)\ignorespaces}%
    \def\contentslineAsubsection##1##2##3{\ignorespaces}%
    \def\contentslineAparagraph##1##2##3{\ignorespaces}%
    \@starttoc{toc}.}}
\makeatother

\makeatletter
\def\Cases#1{\mbox{$
  \left\{\,\vcenter{\let\\=\cr \normalbaselines\m@th
    \ialign{$\vphantom(##\hfil$&\quad##\hfil\crcr#1\crcr}}\right.$}}
\makeatother

\usepackage{cite}
\usepackage{stmaryrd}

\title{ ħ: A Plank for Higher-order Attribute Contraction Schemes \\ \emph{Extended Abstract} }
\author{%
  Cynthia Kop \\
  University of Copenhagen
  \and
  Kristoffer H. Rose \\
  Two Sigma Investments, LP
}

\begin{document}
\maketitle

\begin{abstract}\noindent
  We present and formalize ħ, a core (or ``plank'') calculus that can serve as the foundation
  for several compiler specification languages, notably \CRSX (Combinatory Reductions Systems with
  eXtensions) \HAX (Higher-order Attribute Contraction Schemes), and \textsf{TransScript}.
  We discuss how the ħ typing and formation rules introduce the necessary restrictions to
  ensure that rewriting is well-defined, even in the presence of ħ's powerful extensions for
  manipulating free variables and environments as first class elements (including in pattern
  matching).
\end{abstract}

\section{Introduction}\label{sec:intro}

Several systems that manipulate programs, so-called \emph{meta-programming} systems, have emerged
over the years, ranging from generic specification languages, where the goal is to derive
implementations from formal specifications of the program language semantics, all the way to tools
that support specific aspects of program execution or compiler generation.

One direction has been to combine \emph{higher order rewriting}~\cite{Jouannaud:klop2005} with
\emph{higher order abstract syntax} (HOAS) \cite{PfenningElliot:pldi1988}. This approach is used by
\CRSX (Combinatory Reduction Systems with eXtensions)~\cite{Rose:1996}, developed for writing
industrial compilers at IBM Research~\cite{Rose:rta2011}, and the derived systems \HAX (Higher-order
Attribute Contration Schemes)~\cite{Rose:ts2015}, developed to teach compiler construction at
NYU~\cite{RoseRose:cims2015}, and \textsf{TransScript} \cite{}, developed to support industrial
strength fully automatic compiler generation.  The implementation of the full \CRSX
language~\cite{crsx} turned out to be quite complex, and over time we have developed notions of what
the ``core'' elements of such languages should be.

ħ is our attempt at providing such an extended foundation.  ħ is specifically based on Aczel's
\emph{Contraction Schemes}~\cite{Aczel:1978} (where the restriction we shall see of only permitting
abstractions as constructor arguments is first found) with the substitution notion from Klop's CRSs,
\emph{Combinatory Reduction Systems}~\cite{Klop+:tcs1993} and imposing a typing discipline similar
to the the second-order limitation of Blanqui's \emph{Inductive Data Type
  Systems}~\cite{BlanquiJouannaudOkada:tcs2002}. The extension of ħ is the mechanism used to add
first class \emph{environments} and \emph{free variables} in a formal way, and integrate those with
the type discipline and pattern matching; these features are derived from the \CRSX
system~\cite{Rose:rta2011}, where they were never provided a formal foundation.

\begin{example}[$βη$]
  The canonical higher order rewriting example, λ-calculus with β and η rewrite rules over closed
  terms, is expressed as follows in~ħ:
  \begin{hacs}
  L scheme Lam([L]L);
  L scheme Ap(L,L);
  L rule Ap(Lam([x]#M(x)), #N) →  #M(#N);
  L rule Lam([x]Ap(#M(), x)) →  #M();
  \end{hacs}
  The example shows CRS-style higher order matching, where patterns match all occurrences of bound
  variables, allowing rules to do \emph{substitution} of bound variables.
  In the first rule, "#M(x)" in the left-hand side matches a term containing the bound variable
  x, while in the right-hand side "#M(#N)" achieves the substitution ``"#M[#N/x]"'' of the usual β
  rule. CRS notably also makes it possible to test for the \emph{absence} of binder arguments to
  meta-variables: in the last rule, "#M()" is not the same as "#M(x)", and expresses that "x" can
  \emph{not} occur in subterms matching "#M".
\end{example}

\begin{example}
  Here is a traditional ``call-by-value'' λ calculus evaluator, using HOAS and an environment.
  \begin{hacs}[numbers=right]
  L data Lam([L]L);
  L data Ap(L, L);
  L variable;

  L scheme Eval(L, {L:L});
  L rule Eval(Lam([x]#B(x)), {#env})
    → Lam([x]#B(x));
  L rule Eval(Ap(#F, #A), {#env})
    → Apply(Eval(#F, {#env}),
              Eval(#A, {#env}), {#env});
  L rule Eval(x, {#env; x : #V}) → #V;

  L scheme Apply(L, L, {L:L});
  L rule Apply(Lam([x]#B(x)), #V, {#env})
    → Eval(#B(z), {#env, z : #V}); 
  \end{hacs}
  The "variable" declaration ensures that variables in the "L" sort are syntactic, so they can be
  matched in line 11 (free) and line 13 (bound), where we in the latter substitute it with a fresh
  variable, "z" in line 14, in the term and environment.
\end{example}

\section{The Calculus}
\label{sec:overview}

\begin{definition}[ħ syntax]\label{def:syntax}
  The ħ syntax is summarized as follows (with overbars denoting vectors):
  \begin{align*}
    \tag{ħ{}Script}
    H &::= \ov{D} 
    \\
    \tag{Declaration}
    D &::= S~\kw{data}~d\,\kw(\,\ov{F}\,\kw)\,\kw;\\*[-1pt]
    &~\bigm| S~\kw{scheme}~f\,\kw(\,\ov{F}\,\kw)\,\kw;\\*[-1pt]
    &~\bigm| S~\kw{variable}\,\kw;\\*[-1pt]
    &~\bigm| S~\kw{rule}~T~\kw{$→$}~T\,\kw;
    \\
    \tag{Form}
    F &::= \kw[\,\ov{S}\,\kw]\,S
    \bigm| \kwm\{\, S:S \,\kwm\}
    \\
    \tag{Sort}
    S &::= s\,\kwm{⟨}\,\ov{S}\,\kwm{⟩}
    \bigm| α
    \\[\jot]
    \tag{Term}
    T &::= c\,\kw(\,\ov{P}\,\kw)
    \bigm| v
    \bigm| m\,\kw(\,\ov{T}\,\kw)
    \\
    \tag{Piece}
    P &::= \kw[\,\ov{v}\,\kw]\,T
    \bigm| \kwm\{\, \ov{A} \,\kwm\}
    \\
    \tag{Association}
    A &::= v\,\kw:\,T
    \bigm| \kw{$¬$}\,v\kw:
    \bigm| \kw\,m\,\kw(\,\ov{v}\,\kw)
  \end{align*}
  The top level of a ħ script is $H$ and the grammar assumes that we have three categories of
  identifier terminals defined (with representative exemplars from the examples):
  \begin{itemize}
  \item $c,s,d,f ∈ \mathcal{C}$ stand for \emph{constructor} terminals ("Lam", "Ap").
  \item $v,α ∈ \mathcal{V}$ stand for \emph{variable} terminals ("x", "z").
  \item $m ∈ \mathcal{M}$ stands for \emph{meta-variable} terminals ("#M", "#env").
  \end{itemize}
\end{definition}

The ħ formalism includes traditional ``constructor'' term rewriting systems, where there is a
distinction between \emph{defined} "scheme" symbols and "data" symbols. However, unlike functional
programming, ħ allows for normal forms that include (incompletely) defined symbols.

\section{Sorting}
\label{sec:sorting}

In this Section we define ħ script formation formally by restricting the terms of the grammar in
Definition~\ref{def:syntax} further to only allow well-formed ``sortable'' scripts. Informally,
sorting ensures that
\begin{itemize}
\item pattern and contraction restrictions are obeyed;
\item our special ``syntactic variables'' are used correctly;
\item binders are used in the shape declared for constructors;
\item subterms (including variable and meta-variable occurrences) have the right sort;
\item association keys and values have the proper sorts.
\end{itemize}

\begin{definition}
  A \emph{global sort environment} $Γ$ is a structure that combines
  $Γ_{\op{rank}}\colon \mathcal{C} → \mathcal{N}$: the rank of each sort
  constructor (all sorts must be fully applied, there are no higher kinds);
  $Γ_{\op{hasvar}}$: a set of sort names (the sorts that allow variables);
  $Γ_{\op{con}}\colon \mathcal{C} → S×F^*$: a map from constructor names to pairs of a sort and a list
  of forms (the pair consists of the construction's sort and the shape of the arguments); and
  $Γ_{\op{fun}}$: a set of constructor names (those declared with "scheme").
\end{definition}

\begin{definition}
  A \emph{rule environment} $Δ$ is a structure that combines
  $Δ_{\op{var}}\colon \mathcal{V} → S$: a mapping from variable names to sorts (the sort assigned
  to each variable for the rule) and
  $Δ_{\op{meta}}\colon \mathcal{M} → MF$: a mapping from meta-variable names to ``meta-forms''
  defined by
  \begin{equation}
    MF ::= \ov{S}⇒S \mid \ov{S}⇒\{S{:}S\} \tag{MetaForm}
  \end{equation}
  $MF$ captures the difference between regular meta-variables and ``catch-all'' ones:
  The shape $\ov{S}⇒S$ is used for meta-variables that need to be meta-applied to arguments with the
  sorts $\ov{S}$ to then form a term of the sort~$S$.
  The shape $\ov{S}⇒\{S_1{:}S_2\}$ is used for meta-variables that catch all the associations in an
  association list from $S_1$ to~$S_2$.
\end{definition}

\begin{definition}
  A ħ script $H$ is \emph{well formed} if we can prove $⊢H$ with the rule
  \begin{align}
    &
    \dfrac
    { Γ ⊢ D_1 ~~\cdots~~ Γ ⊢ D_n }
    { ⊢ D_1…D_n }
    \qquad\qquad (∃Γ)
    \tag{SH}
  \end{align}
  using the rules below for sorting each declaration.
\end{definition}

Thus sorting relies on a sort environment ``witness'' to establish that a script is well sorted. In
practice, $Γ$ will be assembled from the constraints of the component declarations. Because all top
level symbols are explicitly sorted, sort assignment is not problematic.

\begin{definition}
  A ħ declaration $D$ is well-sorted for a sort environment $Γ$ if we can prove $Γ⊢D$ with the
  SD-* rules:
  \begin{align}
    &\dfrac
    { Γ ⊢ s⟨\ov{α}⟩ }
    { Γ ⊢ s⟨\ov{α}⟩~\kw{data}~d\kw( \ov{F} \kw)\,\kw; }
    \quad Γ_{\op{con}}(d) = \left⟨ s⟨\ov{α}⟩, \ov{F} \right⟩,~d∉Γ_{\op{fun}}
    \tag{SD-Data}
    \\[1ex]
    &\dfrac
    { Γ ⊢ S }
    { Γ ⊢ S~\kw{scheme}~f\kw( \ov{F} \kw)\,\kw; }
    \quad Γ_{\op{con}}(f) = \left⟨ S, \ov{F} \right⟩,~f∈Γ_{\op{fun}}
    \tag{SD-Fun}
    \\[1ex]
    &\dfrac
    { Γ ⊢ s⟨\ov{α}⟩ }
    { Γ ⊢ s⟨\ov{α}⟩~\kw{variable}\,\kw; }
    \quad s ∈ Γ_{\op{hasvar}}
    \tag{SD-Var}
    \\[1ex]
    &\dfrac
    { Γ ⊢ S \quad Γ,Δ,V,\op{Pat},ε ⊢ T_1 : S  \quad Γ,Δ,V,\op{Con},ε ⊢ T_2 : S }
    { Γ ⊢ S~\kw{rule}~T_1~\kw{→}~T_2\,\kw; }
    \notag
    \\*
    &\qquad\qquad\qquad\qquad\Cases{
      (∃Δ) \\
      V_i = \op{NonAssocVars(T_i)}
    }
    \tag{SD-Rule}
  \end{align}
  Rules \thetag{SD-Data,SD-Fun} express that constructors are recorded correct in the environment;
  note that data constructors must belong to a unique named sort. \thetag{SD-Var} records sorts with
  syntactic varables, and \thetag{SD-Rule} checks the sorts and well-formedness of the terms in
  rules; the ``NonAssocVars'' function returns all variables in a term that are \emph{not} inside an
  association.
\end{definition}

\begin{definition}
  A sort denotation $S$ is well-sorted for a sort environment $Γ$ if we can prove $Γ⊢S$ with the
  SS-* rules:
  \begin{align}
    %
    &\dfrac
    { Γ ⊢ S_1 ~~\cdots~~ Γ ⊢ S_n }
    { Γ ⊢ s⟨S_1,…,S_n⟩ }
    && Γ_{\op{rank}}(s) = n 
    \tag{SS-Cons}
    \\[1ex]
    &\dfrac
    {}
    { Γ ⊢ α }
    \tag{SS-Var}
  \end{align}
  Essentially, sorts are well-formed when they have consistent rank.
\end{definition}

\begin{figure*}[p]\small
  \vspace*{-3em}
  \begin{align}
    \intertext{\shoveright{Sorting of Term\hfil\smash{\fbox{$ Γ,Δ,V,TC,\ov{\mathcal{V}} ⊢ T : S $}}}}
    &\dfrac
    { Γ,Δ,V,\op{InPat},\ov{v} ⊢ P_1 : F_1 ~~\cdots~~ Γ,Δ,V,\op{InPat},\ov{v} ⊢ P_n : F_n }
    { Γ,Δ,V,\op{Pat},\ov{v} ⊢ f\,\kw(\,P_1\kw,…\kw,P_n\,\kw) : S }
    &&\Cases{
      f∈Γ_{\op{fun}}\\
      Γ_{\op{con}}(f) = \left⟨ S, (F_1\kw,…\kw,F_n) \right⟩
    }
    \tag{SMP-Fun}
    \\[1ex]
    &\dfrac
    { Γ,Δ,V,\op{InPat},\ov{v} ⊢ P_1 : F_1 ~~\cdots~~ Γ,Δ,V,\op{InPat},\ov{v} ⊢ P_n : F_n }
    { Γ,Δ,V,\op{InPat},\ov{v} ⊢ d\,\kw(\,P_1\kw,…\kw,P_n\,\kw) : S }
    &&\Cases{
      d∉Γ_{\op{fun}}\\
      Γ_{\op{con}}(d) = \left⟨ S, (F_1\kw,…\kw,F_n) \right⟩
    }
    \tag{SMP-Data}
    \\[1ex]
    &\dfrac
    { }
    { Γ,Δ,V,\op{InPat},\ov{v} ⊢ m\,\kw(\,w_1\kw,…\kw,w_n\,\kw) : S }
    &&\Cases{
      Δ_{\op{meta}}(m) = \left( (S_1,…,S_n)⇒S \right) \\
      ∀i\colon w_i∈\ov{v} \\
      ∀i\colon Δ_{\op{var}}(w_i) = S_i \\
      \text{All the $w_i$ are different}
    }
    \tag{SMP-Meta}
    \\[1ex]
    &\dfrac
    { }
    { Γ,Δ,V,\op{InPat},\ov{v} ⊢ w : s⟨\ov{S}⟩ }
    &&\Cases{
      Δ_{\op{var}}(w) = s⟨\ov{S}⟩ \\
      s ∈ Γ_{\op{hasvar}}
    }
    \tag{SMP-Var}
    \\[1em]
    &\dfrac
    { Γ,Δ,V,\op{Con},\ov{v} ⊢ P_1 : F_1 ~~\cdots~~ Γ,Δ,V,\op{Con},\ov{v} ⊢ P_n : F_n }
    { Γ,Δ,V,\op{Con},\ov{v} ⊢ c\,\kw(\,P_1\kw,…\kw,P_n\,\kw) : S }
    && Γ_{\op{con}}(c) = \left⟨ S, (F_1\kw,…\kw,F_n) \right⟩
    \tag{SMC-Cons}
    \\[1ex]
    &\dfrac
    { Γ,Δ,V,\op{Sub},\ov{v} ⊢ T_1 : S_1 ~~\cdots~~  Γ,Δ,V,\op{Sub},\ov{v} ⊢ T_n : S_n }
    { Γ,Δ,V,\op{Con},\ov{v} ⊢ m\,\kw(\,T_1\kw,…\kw,T_n\,\kw) : S }
    && Δ_{\op{meta}}(m) = \left( (S_1,…,S_n)⇒S \right)
    \tag{SMC-Meta}
    \\[1ex]
    &\dfrac
    { }
    { Γ,Δ,V,\op{Con},\ov{v} ⊢ w : s⟨\ov{S}⟩ }
    &&\Cases{
      Δ_{\op{var}}(w) = s⟨\ov{S}⟩ \\
      s ∈ Γ_{\op{hasvar}}
    }
    \tag{SMC-Var}
    \\[1em]
    &\dfrac
    { Γ,Δ,V,\op{Con},\ov{v} ⊢ c\,\kw(\,P_1\kw,…\kw,P_n\,\kw) : s⟨\ov{S}⟩ }
    { Γ,Δ,V,\op{Sub},\ov{v} ⊢ c\,\kw(\,P_1\kw,…\kw,P_n\,\kw) : s⟨\ov{S}⟩ }
    && s ∉ Γ_{\op{hasvar}}
    \tag{SMS-Cons}
    \\[1ex]
    &\dfrac
    { Γ,Δ,V,\op{Con},\ov{v} ⊢ m\,\kw(\,T_1\kw,…\kw,T_n\,\kw) : s⟨\ov{S}⟩ }
    { Γ,Δ,V,\op{Sub},\ov{v} ⊢ m\,\kw(\,T_1\kw,…\kw,T_n\,\kw) : s⟨\ov{S}⟩ }
    && s ∉ Γ_{\op{hasvar}}
    \tag{SMS-Meta}
    \\[1ex]
    &\dfrac
    { }
    { Γ,Δ,V,\op{Sub},\ov{v} ⊢ w : S }
    && Δ_{\op{var}}(w) = S
    \tag{SMS-Var}
    \\[1ex]
    \intertext{\shoveright{Sorting of Piece\hfil\smash{\fbox{$ Γ,Δ,V,TC,\ov{\mathcal{V}} ⊢ P : F $}}}}
    &\dfrac
    { Γ,Δ∪Δ',V,TC,(\ov{v}\,\ov{w}) ⊢ T : S }
    { Γ,Δ,V,TC,\ov{v} ⊢ \kw[\,\ov{w}\,\kw]\,T : [\ov{S}]S }
    && (∃Δ'), Δ'_{\op{var}}(\ov{w}) = \ov{S}
    \tag{SP-Bind}
    \\[1ex]
    &\dfrac
    { Γ,Δ,V,TC,\ov{v} ⊢ A_1 : \{S{:}S'\} ~~\cdots~~ Γ,Δ,V,TC,\ov{v} ⊢ A_n : \{S{:}S'\} }
    { Γ,Δ,V,TC,\ov{v} ⊢ \kwm\{\,A_1\kw,…\kw,A_n\,\kwm\} : \{S{:}S'\} }
    \tag{SP-Assoc}
    \\[1ex]
    \intertext{\shoveright{Sorting of Association\hfil\smash{\fbox{$ Γ,Δ,V,TC,\ov{\mathcal{V}} ⊢ A : \{S{:}S\} $}}}}
    &\dfrac
    { Γ,Δ,V,TC,\ov{v} ⊢ w : S \quad Γ,Δ,V∪V',TC,\ov{v} ⊢ T' : S' }
    { Γ,Δ,V,TC,\ov{v} ⊢ w\kw:T' : \{S{:}S'\} } 
    \tag{SA-Map}
    &&\Cases{
      w ∈ V \\
      V' = \op{NonAssocVars}(T')
    }
    \\[1ex]
    &\dfrac
    { Γ,Δ,V,\op{InPat},\ov{v} ⊢ w : S }
    { Γ,Δ,V,\op{InPat},\ov{v} ⊢ {\kwm{¬}w\kw{:}} : \{S{:}S'\} }
    \tag{SAP-Not}
    \\[1ex]
    &\dfrac
    { }
    { Γ,Δ,V,\op{InPat},\ov{v} ⊢ m\,\kw(\,w_1\kw,…\kw,w_n\,\kw) : \{S{:}S'\} }
    &&\Cases{
      Δ_{\op{meta}}(m) = \left( (S_1,…,S_n)⇒\{S{:}S'\} \right) \\
      ∀i\colon w_i∈\ov{v} \\
      ∀i\colon Δ_{\op{var}}(w_i) = S_i
    }
    \tag{SAP-All}
    \\[1ex]
    &\dfrac
    { Γ,Δ,V,\op{Sub},\ov{v} ⊢ T_1 : S_1 ~~\cdots~~  Γ,Δ,V,\op{Sub},\ov{v} ⊢ T_n : S_n }
    { Γ,Δ,V,\op{Con},\ov{v} ⊢ m\,\kw(\,T_1\kw,…\kw,T_n\,\kw) : \{S{:}S'\} }
    && Δ_{\op{meta}}(m) = \left( (S_1,…,S_n)⇒\{S{:}S'\} \right)
    \tag{SAC-All}
  \end{align}
  \vspace*{-1em}
  \caption{ħ term sorting rules.}
  \label{fig:termsortrules}
\end{figure*}

\begin{definition}
  Given a sort environment $Γ$, a rule environment $Δ$, and a variable set~$V$. A term $T$ is
  \emph{well-formed of sort $S$} in term context $TC$ with
  \begin{equation}
    TC ::= \op{Pat} \mid \op{InPat} \mid \op{Con} \mid \op{Sub} \tag{TermContext}
  \end{equation}
  and bound variables $\ov{v}$ if we can prove
  \begin{equation*}
    Γ,Δ,V,TC,\ov{v} ⊢ T : S  
  \end{equation*}
  using the rules in Figure~\ref{fig:termsortrules}.
\end{definition}

The rules in Figure~\ref{fig:termsortrules} give the primary term rules per term context $TC$, which
indicates the context of our term fragment:
\begin{itemize}
\item ``Pat'' indicates the pattern (left side) of a rule, at the outermost level.
\item ``InPat'' is used inside a piece of the pattern of a rule, not the outermost level.
\item ``Con'' denotes any location in the contraction (right side) of a rule except a substitution.
\item ``Sub'' denotes a substitution location (an immediate child of a meta-application) in the
  contraction (right side) of a rule.
\end{itemize}
The \thetag{SMP-*} rules handle patterns. The first rule \thetag{SMP-Fun} is the entry rule
for patterns, with $TC=\op{Pat}$, which must be function constructions, and have pieces that are
well-sorted. Fragments of patterns are then handled by the three following rules,
\thetag{SMP-Data,SMP-Meta,SMP-Var}, with $TC=\op{InPat}$, which capture the sort propagation as
well as the special constraints for patterns:
\thetag{SMP-Data} sort checks that only data constructors are allowed;
\thetag{SMP-Meta} verifies that pattern meta-application arguments are restricted to distinct bound
variables;
\thetag{SMP-Var} verifies that other instances of variables only occur where a \kw{variable}
declaration explicitly permits it.

The \thetag{SMC-*} rules handle contraction terms.
\thetag{SMC-Cons} verifies that each construction (function or data) is well sorted.
\thetag{SMC-Meta} verifies that every meta-application is sorted consistently with the rule
environment, and checks that the implied substitutions are well formed.
\thetag{SMC-Var} checks that variables are used at the right sort, and are permitted (by a
\kw{variable} declaration for the sort).

The \thetag{SMS-*} rules handle contraction substitution arguments. They really just wrap the
corresponding \thetag{SMC-*} rules except that \thetag{SMS-Cons,SMS-Meta} checks that the
non-trivial substitution is permitted by the sort \emph{not} having a \kw{variable} declaration,
and \thetag{SMS-Var} just checks the sort of the replacement variable \emph{without} any
constraints on whether the sort has a \kw{variable} declaration.

The \thetag{SP-*} rules handle ``pieces,'' \ie, parameters of constructions (which at the
outermost level have a different structure than other subterms). Note that these do \emph{not}
depend on the term context, which is merely passed to the premises. (From the other rules we can
see that the term context will always be \op{InPat} or \op{Con}.)
\thetag{SP-Bind} handles scopes. It creates a \emph{locally extended} version of the rule
environment, $Δ'$, which extends the variable bindings part of $Δ$ with the binders in the scope;
the sorts of these are fixed by the parent construction.
\thetag{SP-Assoc} handles collections of associations. These are checked by separate individual
rules for association, below.

The \thetag{SA*} rules handle associations. These are categorized in a slightly different ways
than the others above, as associations have different rules in patterns and contractions.
\thetag{SA-Map} gives the rule for a simple mappings. The only unusual requirement is the side
condition that the variable, $w$, must also occur elsewhere in the term as a non-key, a requirement
that ensures that matching is deterministic (in essence avoiding the ``axiom of choice'' for the set
of keys).
The two \thetag{SAP-*} rules express the constraints on the other two forms that can occur in
patterns, and \thetag{SAC-All} that environment copy is unconstrained in cntractions.

\section{Implementation}
\label{sec:imp}

The \plank language has been implemented in the \CRSX project from scratch, and is available from
\url{http://github.com/crsx/core/plank}. At the time of writing, it still has errors, but these are
all minor and we expect to root them out by the time of the workshop.

\section{Conclusion}
\label{sec:conc}

With ħ, we have presented a rather small calculus that can serve as the underlying formalism for
reasoning about, as well as implementing, serious compiler generation languages that support native
higher-order abstract syntax. We have a full implementation.

\paragraph*{Related work.}

We would like to give credit to SIS~\cite{Mosses:daimi1979}, which shares with ħ the use of
\emph{simplification} using a λ-calculus based formalism.

The most prominent system that supports implementation of compilers in formal (rewriting and
attribute grammar) form is ASF+SDF~\cite{Brand+:toplas2002}, which is based on first order
rewriting. While modules have been added for symbol table management, these lack the full
integration and easy way to handle scoped intermediate languages. The successor,
Rascal~\cite{Bos+:eptcs2011} adds a module for HOAS, but Rascal specifications operate in a world of
side effects, which we find hard to reconcile with higher-order term structures (with scopes).

The notion of ``higher-order'' used by ħ is similar to but not quite the same as in higher-order
attribute grammars (HAG)~\cite{VogtSwierstraKuiper:pldi1989}. Like HAGs, ħ specifications permit
constructing and passing of abstract syntax fragments in attributes but the ``higher order'' aspect
of ħ also covers the rewriting side, where we can build parameterized abstractions over any part
of a specification, including with attributes. Indeed, one can use substitution inside attributes,
and have absence of attributes and substitution block rewriting.

\paragraph*{Future work.}

For the full version of this work, we plan to complete the implementation by porting the old \CRSX
pattern matching compiler and possibly looking to integrate with the TransScript ANTLR-based parser
frontend. On the theory side we still owe the world a proper formal development of the properties of
matching in \plank.

\paragraph*{Acknowledgements.}

The authors are grateful to Lionel Villard, Maria Schett, and Julian Nagele, for continuous valuable
discussions in the \CRSX project during the development of this work.
The project is supported by the Marie Sk{\l}odowska-Curie action ``HORIP'', program
H2020-MSCA-IF-2014, 658162.

\bibliography{crs}

\end{document}